\documentclass[aps,pra,twocolumn,showpacs,amsmath,amssymb,nofootinbib,superscriptaddress]{revtex4}

\newcommand{\bra}[1]{\langle#1|}
\newcommand{\ket}[1]{|#1\rangle}

\usepackage[dvips]{graphicx}
\usepackage{mathrsfs}

\begin{document}

\bibliographystyle{apsrev}

\title{Quantum state tomography of single photons in the spectral degree of freedom}

\author{Peter P. Rohde}
\email[]{rohde@physics.uq.edu.au}
\homepage{http://www.physics.uq.edu.au/people/rohde/}
\affiliation{Centre for Quantum Computer Technology, Department of Physics\\ University of Queensland, Brisbane, QLD 4072, Australia}

\date{\today}

\frenchspacing

\begin{abstract}
Quantum State Tomography (QST) of optical states is typically performed in the photon number degree of freedom, a procedure which is well understood and has been experimentally demonstrated. However, optical states have other degrees of freedom than just photon number, such as the spatial and temporal/spectral ones. Full characterization of photonic states requires state reconstruction in these additional degrees of freedom. In this paper we present a technique for performing QST of single photon states in the spectral degree of freedom. This is of importance, for example, in quantum information processing applications, which typically impose strict requirements on the purity and distinguishability of independently produced single photons. The described technique allows for full reconstruction of the spectral density matrix, allowing the purity and distinguishability of different sources to be readily calculated.
\end{abstract}


\maketitle

The preparation of single photon states is vital to the development of many optical quantum information processing protocols \cite{bib:KLM01}. This has motivated much research into photon engineering techniques \cite{bib:URen03,bib:Brunel99,bib:Keller04,bib:Kurtsiefer00,bib:Lounis00,bib:McKeever92,bib:Santori01,bib:Santori02}. In order to be useful for quantum information processing applications there are typically very stringent requirements on the purity, distinguishability \cite{bib:RohdeRalph06,bib:RohdeRalphMunro06}, and mode-structure \cite{bib:RohdeRalph05b} of prepared single photons. Therefore, characterizing single photon sources is of great practical interest.

The most general approach to characterizing quantum states is via Quantum State Tomography (QST) \cite{bib:NielsenChuang00}, which, using many identical copies of a given state, prescribes an approach for complete reconstruction of the density operator from the measurement of experimentally accessible observables. In the context of optical state characterization this is typically performed in the photon number degree of freedom \cite{bib:Banaszek96,bib:Banaszek99,bib:Pregnell02}, most notably using Optical Homodyne Tomography (OHT) \cite{bib:Smithey93}. In this paper we describe an approach for performing QST of single photons in the spectral/temporal degree of freedom. From the reconstructed density operator, important measures, such as the spectral purity and distinguishability of different sources, can be readily calculated. Previous work has examined the issue of experimentally determining the temporal wave-packets of single photons \cite{bib:Legero05}. However, our work is more general in that allows for complete reconstruction of the spectral density operator.

\textbf{Conceptual overview ---}
We begin with a conceptual overview to provide some intuition into our protocol. The main intuitive notion upon which our protocol is based is that the relationship between a photon's wavepacket in conjugate domains (e.g. time and frequency) provides information about the purity of the state in those degrees of freedom, and hence the off-diagonal elements of the density matrix. For example, in the limit of pure photons we approach `transform limited' wavepackets, meaning that the frequency and time domain wavepackets are directly related by Fourier transformation. In this case the state is of the form
\begin{equation}
\ket\psi = \int \psi(\omega)\hat{a}^\dag(\omega)\,\mathrm{d}\omega \ket{vac} =\int \tilde\psi(t)\tilde{a}^\dag(t)\,\mathrm{d}t \ket{vac},
\end{equation}
where $\hat{a}^\dag(\omega)$ ($\tilde{a}^\dag(t)$) is the single frequency (time) photonic creation operator, $\tilde\psi(t)$ is the inverse Fourier transform of $\psi(\omega)$, and the integral run over all frequencies (times). For impure states this relationship breaks down. As an example, consider a photon which is a mixture of identical, but temporally displaced wavepackets. This might arise when a photon source exhibits `time-jitter'. Such a photon can be described in the form
\begin{equation}
\hat\rho=\int\!\!\int\!\!\int \tilde{f}(t_c)\tilde\psi(t_1)\tilde\psi^*(t_2) \ket{t_1-t_c}\bra{t_2-t_c}\,\mathrm{d}t_1\,\mathrm{d}t_2\, \mathrm{d}t_c,
\end{equation}
where $\tilde{f}(t)$ characterizes the mixture (i.e. time-jitter), $\tilde\psi(t)$ characterizes the temporal wave-packet of the components in the mixture, $t_c$ denotes a particular center time, and $\ket{t}=\tilde{a}^\dag(t)\ket{vac}$ is the time-specific single photon state at time $t$. Measuring this state in the time domain using time-resolving detectors we will observe the envelope of the entire mixture, roughly $\tilde{f}(t)$, assuming $\tilde\psi(t)$ is narrow compared to $\tilde{f}(t)$. However, measuring this state in the frequency domain with frequency-resolving detectors will not yield an envelope given by $\tilde{f}(\omega)$. This is because the Fourier transforms of each of the components in the mixture are identical up to irrelevant phase factors. Thus, in the spectral domain we observe the envelope of the identical pure components of the mixture, i.e. $\psi(\omega)$.

This observation provides a conceptual hint that in order to measure the spectral density matrix we will need to perform transformations in the conjugate domains, followed by a comparison of these transformed states. While this is a very qualitative description, our protocol is in fact a direct implementation of this idea -- we create a superposition of an incident state across two paths, in one path implementing a frequency domain transformation, and in the other a time domain transformation, and finally we implement a `comparison' by interfering the two paths.

\textbf{Description of the protocol ---}
We now describe the protocol in detail. We begin with a single photon state with completely arbitrary density operator in the spectral degree of freedom. This can be expressed in the form
\begin{eqnarray}
\hat\rho_\mathrm{in}&=&\int\!\!\!\int \rho(\omega_1,\omega_2)\hat{a}^\dag(\omega_1)\ket{vac}\bra{vac}\hat{a}(\omega_2)\,\mathrm{d}\omega_1\,\mathrm{d}\omega_2\nonumber\\
&=&\int\!\!\!\int \rho(\omega_1,\omega_2) \ket{\omega_1}\bra{\omega_2}\,\mathrm{d}\omega_1\,\mathrm{d}\omega_2.
\end{eqnarray}
Our goal is to experimentally sample arbitrary elements of the spectral density matrix $\rho(\omega_1,\omega_2)$.

\begin{figure}[!htb]
\includegraphics[width=0.9\columnwidth]{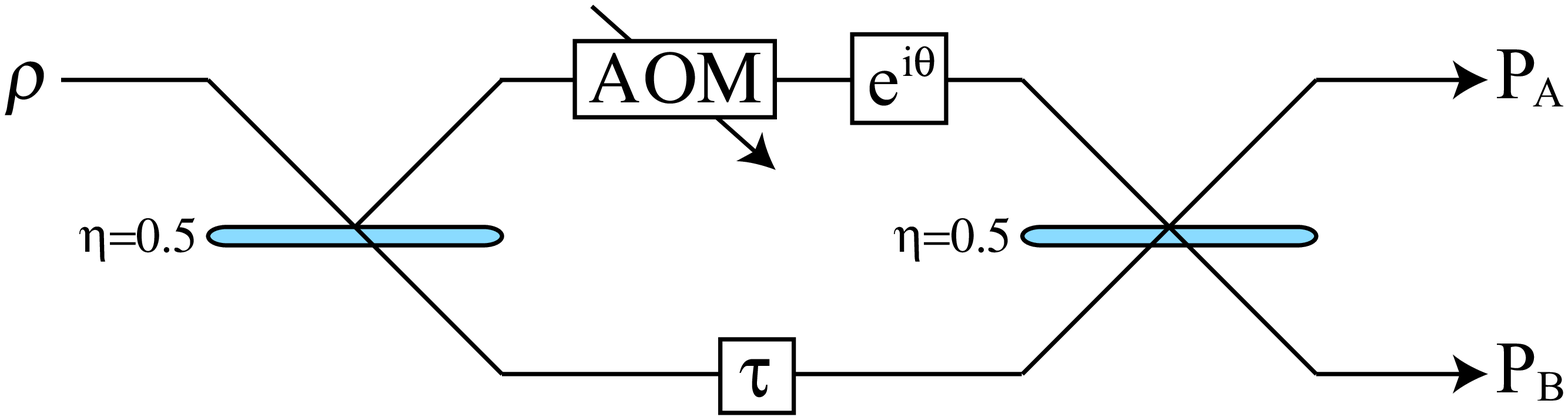}
\caption{The proposed experimental setup. We employ a balanced Mach-Zehnder interferometer. In the upper arm is an Acousto-Optic-Modulator (AOM), followed by a phase shifter. In the lower arm is a time delay. We post-select upon measurements where a photon is detected at one of the outputs.} \label{fig:mach_zehnder}
\end{figure}

We employ a Mach-Zehnder interferometer and perform different temporal and spectral operations in the two arms of the interferometer. The proposed layout is shown in Fig.~\ref{fig:mach_zehnder}.
In the upper arm we induce a tunable spectral shift using an Acousto-Optic-Modulator (AOM). The AOM implements the transformation \cite{bib:Huntington04}
\begin{eqnarray}
\hat{a}_{1,\mathrm{out}}(\omega)&=&\sqrt{1-\xi}\hat{a}_2(\omega)+i\sqrt\xi\hat{a}_1(\omega-\delta)\nonumber\\
\hat{a}_{2,\mathrm{out}}(\omega)&=&\sqrt{1-\xi}\hat{a}_1(\omega)+i\sqrt\xi\hat{a}_2(\omega+\delta),
\end{eqnarray}
where $\delta$ is the modulation frequency applied to the AOM, and $\xi$ is related to the modulation frequency. $\hat{a}_1$ and $\hat{a}_2$ denote distinct spatial modes. We assume the upper arm of the interferometer runs through mode 1, while the vacuum state is incident upon $\hat{a}_2$, and $\hat{a}_{2,\mathrm{out}}$ is discarded. Following the AOM is a phase shifter, which implements the transformation
\begin{equation}
\hat{a}(\omega)\to e^{i\theta}\hat{a}(\omega).
\end{equation}
In the lower arm is a temporal delay, implementing
\begin{equation}
\hat{b}(\omega)\to e^{-i\omega\tau}\hat{b}(\omega).
\end{equation}
The 50/50 beamsplitters are modeled by the transformation
\begin{eqnarray}
\hat{a}_\mathrm{out}(\omega)&=&\left[\hat{a}(\omega)+\hat{b}(\omega)\right]/\sqrt{2},\nonumber\\
\hat{b}_\mathrm{out}(\omega)&=&\left[\hat{a}(\omega)-\hat{b}(\omega)\right]/\sqrt{2}.
\end{eqnarray}
A simple calculation now shows that the conditional output state for detecting a photon in mode $A$ can be expressed in the form
\begin{eqnarray}
\hat\rho_A(\tau,\delta,\theta)&=&\frac{1}{4} \int\!\!\int [e^{-i\tau(\omega_1-\omega_2)} \rho(\omega_1,\omega_2)\nonumber\\
&+& e^{i\theta-i\tau\omega_1} \rho(\omega_1,\omega_2-\delta)\nonumber\\
&+& e^{-i\theta+i\tau\omega_2} \rho(\omega_1-\delta,\omega_2)\nonumber\\
&+& \rho(\omega_1-\delta,\omega_2-\delta)] \ket{\omega_1}\bra{\omega_2}\,\mathrm{d}\omega_1\,\mathrm{d}\omega_2.
\end{eqnarray}
The probability of detecting a photon at output mode $A$ is just the normalization of this state. Thus,
\begin{eqnarray}
P_A(\tau,\delta,\theta)&=&\mathrm{tr}(\hat\rho_A)\nonumber\\
&=& \frac{1}{4} \int \rho(\omega,\omega) + e^{i\theta-i\tau\omega} \rho(\omega,\omega-\delta)\nonumber\\
&+& e^{-i\theta+i\tau\omega} \rho(\omega-\delta,\omega) + \rho(\omega-\delta,\omega-\delta)\,\mathrm{d}\omega,\nonumber\\
\end{eqnarray}
where we have used the identity $\mathrm{tr}(\ket{\omega_1}\bra{\omega_2}) = \langle\omega_1|\omega_2\rangle = \delta(\omega_1-\omega_2)$. Now we can apply several identities to simplify this expression,
\begin{eqnarray}
&&\int\rho(\omega,\omega)\,\mathrm{d}\omega=\int\rho(\omega-\delta,\omega-\delta)\,\mathrm{d}\omega=\mathrm{tr}(\hat\rho_\mathrm{in})=1,\nonumber\\
&&\int e^{i\tau\omega} \rho(\omega-\delta,\omega)\,\mathrm{d}\omega=\mathcal{F}^{-1}_{\omega\to\tau}[\rho(\omega-\delta,\omega)],\nonumber\\
&&\rho(\omega-\delta,\omega)=\rho(\omega,\omega-\delta)^*,
\end{eqnarray}
where $\mathcal{F}^{-1}$ denotes an inverse Fourier transform and subscripts are used to explicitly denote the change of variable taking place. Thus, we have
\begin{eqnarray} \label{eq:P}
P_A(\tau,\delta,\theta)&=&\frac{1}{2}+\frac{1}{2}\mathrm{Re}\left[e^{i\theta}\mathcal{F}^{-1}_{\omega\to\tau}[\rho(\omega,\omega-\delta)]\right],\nonumber\\
P_B(\tau,\delta,\theta)&=&\frac{1}{2}-\frac{1}{2}\mathrm{Re}\left[e^{i\theta}\mathcal{F}^{-1}_{\omega\to\tau}[\rho(\omega,\omega-\delta)]\right].
\end{eqnarray}
Note that in this derivation we have assumed $\xi=1$, corresponding to no loss into the discarded mode in the AOM. When this is not the case this effectively unbalances the interferometer. This can easily be compensated for by introducing an equal loss into the other arm of the interferometer. While this reduces the overall count rate, it does not undermine the operation of the protocol since we can simply post-select upon events where a photon arrives at one of the output ports. Thus, in our calculations we are justified in restricting ourselves to the lossless case, and Eq.~\ref{eq:P} refers to the post-selected scenario.

\begin{figure*}[!htb]
\includegraphics[width=0.85\textwidth]{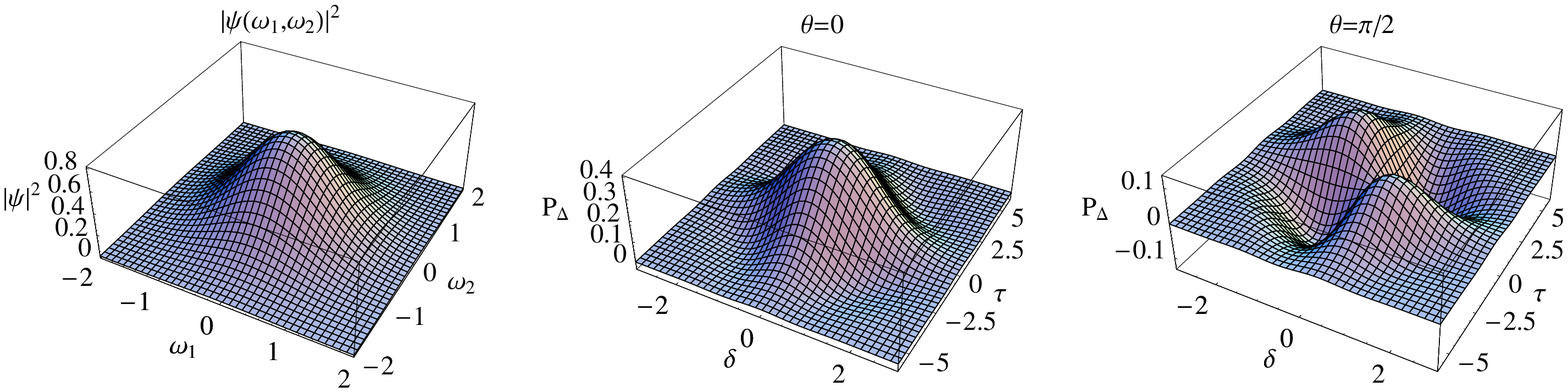}
\includegraphics[width=0.85\textwidth]{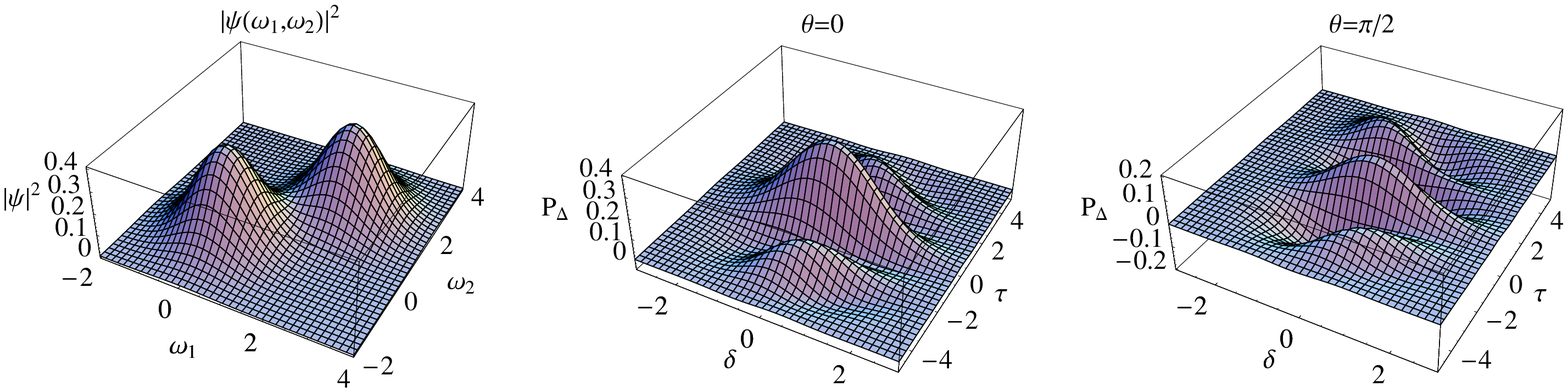}
\includegraphics[width=0.85\textwidth]{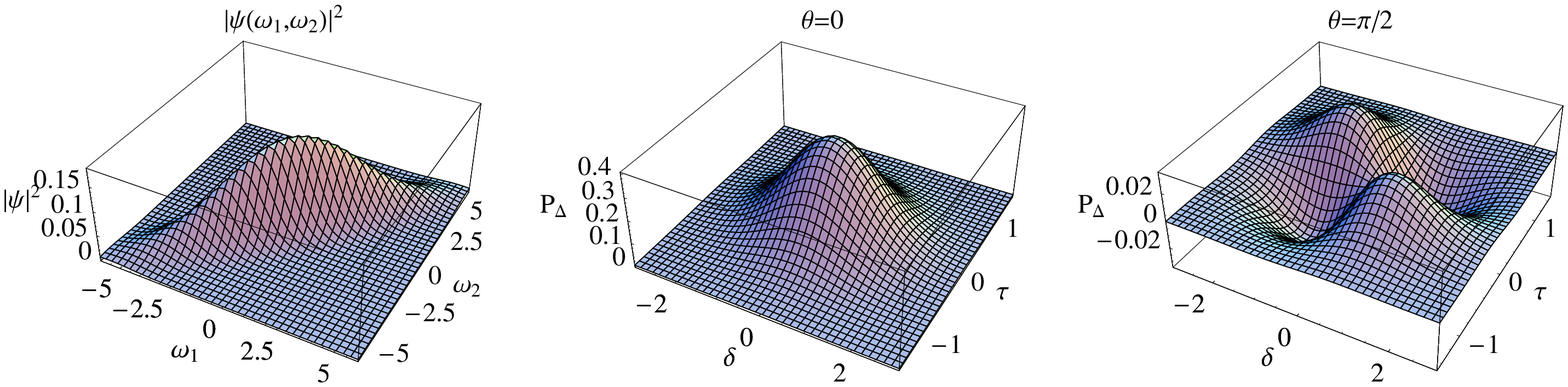}
\caption{Simulated examples of the spectral density operator and corresponding measurement statistics, where $P_\Delta=P_A-P_B$, for (top) a pure Gaussian distributed state, (center) a mixture of two Gaussian distributed states with different center frequencies, and, (bottom) a continuous mixture of Gaussian distributed states with different center frequencies (i.e. frequency jitter).} \label{fig:examples}
\end{figure*}

The procedure for reconstructing the input density matrix is as follows. For every setting of $\tau$ and $\delta$ we perform two measurements, with $\theta=0$ and $\theta=\pi/2$, allowing both the real and imaginary components of $\mathcal{F}_{\omega\to\tau}[\rho(\omega,\omega-\delta)]$ to be measured at every point. This function can then be inverse transformed to establish the diagonal cross section of the input density matrix, $\rho(\omega,\omega-\delta)\,\,\forall\,\,\omega$. By measuring such cross sections for different $\delta$, the complete density matrix can be reconstructed. Several examples of spectral density matrices and their corresponding measurement statistics are illustrated in Fig.~\ref{fig:examples}.

Note that in the case where we do not introduce a frequency shift, $\delta=0$, only the diagonal elements of the density matrix are accessible. This allows the spectral distribution to be measured, but not the coherences between different spectral components. This is expected, and essentially corresponds to existing interferometric techniques for characterizing photon wave-packets.

\textbf{Tolerance against experimental imperfections ---}
We now turn our attention to major sources of experimental imperfection that are likely to arise in the implementation of this scheme. Firstly, our protocol is inherently resilient against source and detector inefficiency. This is because the protocol is post-selected upon events where a photon is detected at one of the outputs.

The second major source of error is likely to be spatial mode-mismatch at the second beamsplitter in the interferometer. We now consider the effect this has on the operation of the protocol. We model this by first expanding our representation for the single photon state to include the transverse spatial degrees of freedom. We define new mode-creation operators $\hat{A}^\dag_{\psi(x,y)}(\omega)$ and $\hat{B}^\dag_{\psi(x,y)}(\omega)$, where $\psi(x,y)$ is the transverse spatial wavefunction of the incident photon. These operators create photons at a particular frequency, but with arbitrary spatial distribution. Formally,
\begin{equation}
\hat{A}^\dag_\psi(\omega)=\int\!\!\int \psi(x,y)\hat{a}^\dag(x,y,\omega)\,\mathrm{d}x\,\mathrm{d}y,
\end{equation}
where $\hat{a}^\dag(x,y,\omega)$ is the space- and frequency-specific creation operator for mode $A$. Note that with this definition we assume the spatial distribution function to be frequency independent. We model mode-mismatch by applying an arbitrary transformation to the spatial wavefunction in one arm of the interferometer, $\hat{A}_\psi(\omega)\to\hat{A}_{\psi'}(\omega)$. The calculation proceeds as before, and we employ the identity
\begin{eqnarray}
&&\mathrm{tr}\left[\hat{A}^\dag_\psi(\omega_1)\ket{vac}\bra{vac}\hat{A}_{\psi'}(\omega_2)\right]\nonumber\\
&&=\delta(\omega_1-\omega_2)\int\!\!\int \psi(x,y)^*\psi'(x,y)\,\mathrm{d}x\,\mathrm{d}y\nonumber\\
&&=\gamma\delta(\omega_1-\omega_2),
\end{eqnarray}
where $\gamma$ parameterizes the degree mode-mismatch ($0\leq\gamma\leq1$, with $\gamma=1$ corresponding to perfect spatial mode overlap, and $\gamma=0$ to no mode overlap). It can now be shown that the output probabilities are given by
\begin{eqnarray} \label{eq:P_mm}
P_A(\tau,\delta,\theta)&=&\frac{1}{2}+\frac{1}{2}\mathrm{Re}\left[\gamma e^{i\theta}\mathcal{F}^{-1}_{\omega\to\tau}[\rho(\omega,\omega-\delta)]\right],\nonumber\\
P_B(\tau,\delta,\theta)&=&\frac{1}{2}-\frac{1}{2}\mathrm{Re}\left[\gamma e^{i\theta}\mathcal{F}^{-1}_{\omega\to\tau}[\rho(\omega,\omega-\delta)]\right].
\end{eqnarray}
Thus, spatial mode-mismatch simply reduces the visibility of the probability fringes by a constant factor. This can be compensated for using some a priori knowledge of the behavior of $P_A$ and $P_B$. When $\tau=\delta=\theta=0$ we expect all photons to exit through port $A$, as per an ordinary balanced Mach-Zehnder interferometer. Thus, by measuring the statistics of $P_A(0,0,0)$ and $P_B(0,0,0)$, the mode-mismatch parameter, $\gamma$, can be directly inferred. With knowledge of $\gamma$, the inversion procedure will faithfully reproduce $\hat\rho_\mathrm{in}$.

\textbf{Variations of the scheme ---}
There is nothing unique about our choice of transformations. Other combinations of transformations in the interferometer arms that are known to work include spectral-shift/spectral-filter (upper/lower arms), temporal-filter/spectral-filter, and temporal-delay/temporal-filter. All of these combinations allow for complete characterization of the spectral density operator in principle. However, due to the difficulty in implementing sub-wavepacket filtering (especially in the time domain), these variations are unlikely to be experimentally feasible.

\textbf{Experimental considerations ---}
In order to successfully reconstruct the spectral density operator we need to be able to induce incremental spectral/temporal shifts over a range on the order of the spectral/temporal bandwidth of the photon. While implementing temporal delays with such high resolution and over an essentially arbitrary scale is experimentally feasible, our ability to induce frequency shifts via AOM's is more limited. Current AOM's can induce frequency shifts on the order of GHz, implying that this technique is limited to photons with, at most, spectral bandwidths of this order. Unfortunately this rules out application of this technique to some widely used single photon engineering techniques, such as ultra-fast parametric down conversion, which inherently have spectral bandwidths outside this range.

\textbf{Conclusion ---}
We have described an approach for tomographically reconstructing the density operator of single photon sources in the spectral degree of freedom. This could prove useful in fully characterizing single photon sources, which is of interest in present quantum information processing applications where the distinguishability and purity of single photons is of great importance.

\begin{acknowledgments}
We thank Timothy Ralph, Christine Silberhorn and Wolfgang Mauerer for helpful discussions. This work was supported by the Australian Research Council and Queensland State Government. We acknowledge partial support by the DTO-funded U.S. Army Research Office Contract No. W911NF-05-0397.
\end{acknowledgments}

\bibliography{paper}

\begin{thebibliography}{19}
\expandafter\ifx\csname natexlab\endcsname\relax\def\natexlab#1{#1}\fi
\expandafter\ifx\csname bibnamefont\endcsname\relax
  \def\bibnamefont#1{#1}\fi
\expandafter\ifx\csname bibfnamefont\endcsname\relax
  \def\bibfnamefont#1{#1}\fi
\expandafter\ifx\csname citenamefont\endcsname\relax
  \def\citenamefont#1{#1}\fi
\expandafter\ifx\csname url\endcsname\relax
  \def\url#1{\texttt{#1}}\fi
\expandafter\ifx\csname urlprefix\endcsname\relax\def\urlprefix{URL }\fi
\providecommand{\bibinfo}[2]{#2}
\providecommand{\eprint}[2][]{\url{#2}}

\bibitem[{\citenamefont{Knill et~al.}(2001)\citenamefont{Knill, Laflamme, and
  Milburn}}]{bib:KLM01}
\bibinfo{author}{\bibfnamefont{E.}~\bibnamefont{Knill}},
  \bibinfo{author}{\bibfnamefont{R.}~\bibnamefont{Laflamme}}, \bibnamefont{and}
  \bibinfo{author}{\bibfnamefont{G.}~\bibnamefont{Milburn}},
  \bibinfo{journal}{Nature (London)} \textbf{\bibinfo{volume}{409}},
  \bibinfo{pages}{46} (\bibinfo{year}{2001}).

\bibitem[{\citenamefont{U'Ren et~al.}(2003)\citenamefont{U'Ren, Banaszek, and
  Walmsley}}]{bib:URen03}
\bibinfo{author}{\bibfnamefont{A.~B.} \bibnamefont{U'Ren}},
  \bibinfo{author}{\bibfnamefont{K.}~\bibnamefont{Banaszek}}, \bibnamefont{and}
  \bibinfo{author}{\bibfnamefont{I.~A.} \bibnamefont{Walmsley}},
  \bibinfo{journal}{Quant. Inf. Comp.} \textbf{\bibinfo{volume}{3}},
  \bibinfo{pages}{480} (\bibinfo{year}{2003}).

\bibitem[{\citenamefont{Brunel et~al.}(1999)\citenamefont{Brunel, Lounis,
  Tamarat, and Orrit}}]{bib:Brunel99}
\bibinfo{author}{\bibfnamefont{C.}~\bibnamefont{Brunel}},
  \bibinfo{author}{\bibfnamefont{B.}~\bibnamefont{Lounis}},
  \bibinfo{author}{\bibfnamefont{P.}~\bibnamefont{Tamarat}}, \bibnamefont{and}
  \bibinfo{author}{\bibfnamefont{M.}~\bibnamefont{Orrit}},
  \bibinfo{journal}{Phys. Rev. Lett.} \textbf{\bibinfo{volume}{83}},
  \bibinfo{pages}{2722} (\bibinfo{year}{1999}).

\bibitem[{\citenamefont{Keller et~al.}(2004)\citenamefont{Keller, Lange,
  Hayasaka, Lange, and Walther}}]{bib:Keller04}
\bibinfo{author}{\bibfnamefont{M.}~\bibnamefont{Keller}},
  \bibinfo{author}{\bibfnamefont{B.}~\bibnamefont{Lange}},
  \bibinfo{author}{\bibfnamefont{K.}~\bibnamefont{Hayasaka}},
  \bibinfo{author}{\bibfnamefont{W.}~\bibnamefont{Lange}}, \bibnamefont{and}
  \bibinfo{author}{\bibfnamefont{H.}~\bibnamefont{Walther}},
  \bibinfo{journal}{Nature (London)} \textbf{\bibinfo{volume}{431}},
  \bibinfo{pages}{1075} (\bibinfo{year}{2004}).

\bibitem[{\citenamefont{Kurtsiefer et~al.}(2000)\citenamefont{Kurtsiefer,
  Mayer, Zarda, and Weinfurter}}]{bib:Kurtsiefer00}
\bibinfo{author}{\bibfnamefont{C.}~\bibnamefont{Kurtsiefer}},
  \bibinfo{author}{\bibfnamefont{S.}~\bibnamefont{Mayer}},
  \bibinfo{author}{\bibfnamefont{P.}~\bibnamefont{Zarda}}, \bibnamefont{and}
  \bibinfo{author}{\bibfnamefont{H.}~\bibnamefont{Weinfurter}},
  \bibinfo{journal}{Phys. Rev. Lett.} \textbf{\bibinfo{volume}{89}},
  \bibinfo{pages}{290} (\bibinfo{year}{2000}).

\bibitem[{\citenamefont{Lounis and Moerner}(2000)}]{bib:Lounis00}
\bibinfo{author}{\bibfnamefont{B.}~\bibnamefont{Lounis}} \bibnamefont{and}
  \bibinfo{author}{\bibfnamefont{W.~E.} \bibnamefont{Moerner}},
  \bibinfo{journal}{Nature (London)} \textbf{\bibinfo{volume}{407}},
  \bibinfo{pages}{491} (\bibinfo{year}{2000}).

\bibitem[{\citenamefont{McKeever et~al.}(2004)\citenamefont{McKeever, Boca,
  Boozer, Miller, Buck, Kuzmich, and Kimble}}]{bib:McKeever92}
\bibinfo{author}{\bibfnamefont{J.}~\bibnamefont{McKeever}},
  \bibinfo{author}{\bibfnamefont{A.}~\bibnamefont{Boca}},
  \bibinfo{author}{\bibfnamefont{A.~D.} \bibnamefont{Boozer}},
  \bibinfo{author}{\bibfnamefont{R.}~\bibnamefont{Miller}},
  \bibinfo{author}{\bibfnamefont{J.~R.} \bibnamefont{Buck}},
  \bibinfo{author}{\bibfnamefont{A.}~\bibnamefont{Kuzmich}}, \bibnamefont{and}
  \bibinfo{author}{\bibfnamefont{H.~J.} \bibnamefont{Kimble}},
  \bibinfo{journal}{Science} \textbf{\bibinfo{volume}{303}},
  \bibinfo{pages}{1992} (\bibinfo{year}{2004}).

\bibitem[{\citenamefont{Santori et~al.}(2001)\citenamefont{Santori, Pelton,
  Solomon, Dale, and Yamamoto}}]{bib:Santori01}
\bibinfo{author}{\bibfnamefont{C.}~\bibnamefont{Santori}},
  \bibinfo{author}{\bibfnamefont{M.}~\bibnamefont{Pelton}},
  \bibinfo{author}{\bibfnamefont{G.}~\bibnamefont{Solomon}},
  \bibinfo{author}{\bibfnamefont{Y.}~\bibnamefont{Dale}}, \bibnamefont{and}
  \bibinfo{author}{\bibfnamefont{Y.}~\bibnamefont{Yamamoto}},
  \bibinfo{journal}{Phys. Rev. Lett.} \textbf{\bibinfo{volume}{86}},
  \bibinfo{pages}{1502} (\bibinfo{year}{2001}).

\bibitem[{\citenamefont{Santori et~al.}(2002)\citenamefont{Santori, Fattal,
  Vuckovic, Solomon, and Yamamoto}}]{bib:Santori02}
\bibinfo{author}{\bibfnamefont{C.}~\bibnamefont{Santori}},
  \bibinfo{author}{\bibfnamefont{D.}~\bibnamefont{Fattal}},
  \bibinfo{author}{\bibfnamefont{J.}~\bibnamefont{Vuckovic}},
  \bibinfo{author}{\bibfnamefont{G.~S.} \bibnamefont{Solomon}},
  \bibnamefont{and} \bibinfo{author}{\bibfnamefont{Y.}~\bibnamefont{Yamamoto}},
  \bibinfo{journal}{Nature (London)} \textbf{\bibinfo{volume}{419}},
  \bibinfo{pages}{594} (\bibinfo{year}{2002}).

\bibitem[{\citenamefont{Rohde and Ralph}(2006)}]{bib:RohdeRalph06}
\bibinfo{author}{\bibfnamefont{P.~P.} \bibnamefont{Rohde}} \bibnamefont{and}
  \bibinfo{author}{\bibfnamefont{T.~C.} \bibnamefont{Ralph}},
  \bibinfo{journal}{Phys. Rev. A} \textbf{\bibinfo{volume}{73}},
  \bibinfo{pages}{062312} (\bibinfo{year}{2006}).

\bibitem[{\citenamefont{Rohde et~al.}(2006)\citenamefont{Rohde, Ralph, and
  Munro}}]{bib:RohdeRalphMunro06}
\bibinfo{author}{\bibfnamefont{P.~P.} \bibnamefont{Rohde}},
  \bibinfo{author}{\bibfnamefont{T.~C.} \bibnamefont{Ralph}}, \bibnamefont{and}
  \bibinfo{author}{\bibfnamefont{W.~J.} \bibnamefont{Munro}},
  \bibinfo{journal}{Phys. Rev. A} \textbf{\bibinfo{volume}{73}},
  \bibinfo{pages}{030301(R)} (\bibinfo{year}{2006}).

\bibitem[{\citenamefont{Rohde et~al.}(2005)\citenamefont{Rohde, Ralph, and
  Nielsen}}]{bib:RohdeRalph05b}
\bibinfo{author}{\bibfnamefont{P.~P.} \bibnamefont{Rohde}},
  \bibinfo{author}{\bibfnamefont{T.~C.} \bibnamefont{Ralph}}, \bibnamefont{and}
  \bibinfo{author}{\bibfnamefont{M.~A.} \bibnamefont{Nielsen}},
  \bibinfo{journal}{Phys. Rev. A} \textbf{\bibinfo{volume}{72}},
  \bibinfo{pages}{052332} (\bibinfo{year}{2005}).

\bibitem[{\citenamefont{Nielsen and Chuang}(2000)}]{bib:NielsenChuang00}
\bibinfo{author}{\bibfnamefont{M.~A.} \bibnamefont{Nielsen}} \bibnamefont{and}
  \bibinfo{author}{\bibfnamefont{I.~L.} \bibnamefont{Chuang}},
  \emph{\bibinfo{title}{Quantum Computation and Quantum Information}}
  (\bibinfo{publisher}{Cambridge University Press, Cambridge},
  \bibinfo{year}{2000}).

\bibitem[{\citenamefont{Banaszek and Wodkiewicz}(1996)}]{bib:Banaszek96}
\bibinfo{author}{\bibfnamefont{K.}~\bibnamefont{Banaszek}} \bibnamefont{and}
  \bibinfo{author}{\bibfnamefont{K.}~\bibnamefont{Wodkiewicz}},
  \bibinfo{journal}{Phys. Rev. Lett.} \textbf{\bibinfo{volume}{76}},
  \bibinfo{pages}{4344} (\bibinfo{year}{1996}).

\bibitem[{\citenamefont{Banaszek et~al.}(1999)\citenamefont{Banaszek,
  Radzewicz, , W{\' o}dkiewicz, and Krasin{\` s}ki}}]{bib:Banaszek99}
\bibinfo{author}{\bibfnamefont{K.}~\bibnamefont{Banaszek}},
  \bibinfo{author}{\bibfnamefont{C.}~\bibnamefont{Radzewicz}}, ,
  \bibinfo{author}{\bibfnamefont{K.}~\bibnamefont{W{\' o}dkiewicz}},
  \bibnamefont{and} \bibinfo{author}{\bibfnamefont{J.~S.}
  \bibnamefont{Krasin{\` s}ki}}, \bibinfo{journal}{Phys. Rev. A}
  \textbf{\bibinfo{volume}{60}}, \bibinfo{pages}{674} (\bibinfo{year}{1999}).

\bibitem[{\citenamefont{Pregnell and Pegg}(2002)}]{bib:Pregnell02}
\bibinfo{author}{\bibfnamefont{K.~L.} \bibnamefont{Pregnell}} \bibnamefont{and}
  \bibinfo{author}{\bibfnamefont{D.~T.} \bibnamefont{Pegg}},
  \bibinfo{journal}{Phys. Rev. A} \textbf{\bibinfo{volume}{66}},
  \bibinfo{pages}{013810} (\bibinfo{year}{2002}).

\bibitem[{\citenamefont{Smithey et~al.}(1993)\citenamefont{Smithey, Beck,
  Raymer, and Faridani}}]{bib:Smithey93}
\bibinfo{author}{\bibfnamefont{D.~T.} \bibnamefont{Smithey}},
  \bibinfo{author}{\bibfnamefont{M.}~\bibnamefont{Beck}},
  \bibinfo{author}{\bibfnamefont{M.~G.} \bibnamefont{Raymer}},
  \bibnamefont{and} \bibinfo{author}{\bibfnamefont{A.}~\bibnamefont{Faridani}},
  \bibinfo{journal}{Phys. Rev. Lett.} \textbf{\bibinfo{volume}{70}},
  \bibinfo{pages}{1244} (\bibinfo{year}{1993}).

\bibitem[{\citenamefont{Legero et~al.}(2005)\citenamefont{Legero, Wilk, Kuhn,
  and Rempe}}]{bib:Legero05}
\bibinfo{author}{\bibfnamefont{T.}~\bibnamefont{Legero}},
  \bibinfo{author}{\bibfnamefont{T.}~\bibnamefont{Wilk}},
  \bibinfo{author}{\bibfnamefont{A.}~\bibnamefont{Kuhn}}, \bibnamefont{and}
  \bibinfo{author}{\bibfnamefont{G.}~\bibnamefont{Rempe}}
  (\bibinfo{year}{2005}), \eprint{quant-ph/0512023}.

\bibitem[{\citenamefont{Huntington and Ralph}(2004)}]{bib:Huntington04}
\bibinfo{author}{\bibfnamefont{E.~H.} \bibnamefont{Huntington}}
  \bibnamefont{and} \bibinfo{author}{\bibfnamefont{T.~C.} \bibnamefont{Ralph}},
  \bibinfo{journal}{Phys. Rev. A} \textbf{\bibinfo{volume}{69}},
  \bibinfo{pages}{042318} (\bibinfo{year}{2004}).

\end{thebibliography}

\end{document}